\documentstyle[aps]{revtex}

\begin{document}

\input epsf
\newcommand{\infig}[2]{\begin{center}\mbox{ \epsfxsize #1
                       \epsfbox{#2}}\end{center}}
\newcommand{\infigtwo}[3]{\begin{center}\mbox{ \epsfxsize #1
                        \epsfbox{#2} \quad \epsfxsize #1
                        \epsfbox{#3}} \end{center}}
\newcommand{\infigtwon}[4]{\begin{center}\mbox{ \epsfxsize #1
                        \epsfbox{#2} \quad \epsfxsize #3
                        \epsfbox{#4}} \end{center}}

\newcommand{\be}{\begin{equation}}
\newcommand{\ee}{\end{equation}}
\newcommand{\bea}{\begin{eqnarray}}
\newcommand{\eea}{\end{eqnarray}}

\draft

\title{Semiclassical theory of cavity-assisted atom cooling}
\author{Peter Domokos\thanks{On leave from the Research Institute for
Solid State Physics and Optics, Hungarian Academy of Sciences}, Peter
Horak, and Helmut Ritsch}

\address{Institut f{\"u}r Theoretische Physik, Universit{\"a}t Innsbruck, 
Technikerstr.\ 25, A-6020 Innsbruck, Austria.} 
\date{\today}

\maketitle 
 
\begin{abstract} 
We present a systematic semiclassical model for the simulation of the dynamics
of a single two-level atom strongly coupled to a driven high-finesse optical
cavity. From the Fokker-Planck equation of the combined atom-field
Wigner function we derive stochastic differential equations for the atomic
motion and the cavity field. The corresponding noise sources exhibit strong
correlations between the atomic momentum fluctuations and the noise in the
phase quadrature of the cavity field. The model provides an effective tool to
investigate localisation effects as well as cooling and trapping times. In
addition, we can continuously study the transition from a few photon quantum
field to the classical limit of a large coherent field amplitude.
\end{abstract} 
 
%\pacs{PACS numbers: 42.50.Lc, 32.80.Pj, 42.50.Vk} 

%\narrowtext

\section{Introduction}

The center-of-mass motion of atoms strongly coupled to the electromagnetic
field in a microresonator has become a central problem in current research in
cavity quantum electrodynaqmics (QED) \cite{CQED}. Very recently trapping
of a single atom by a single photon in a high-Q optical cavity has been
predicted\cite{horak97} and demonstrated \cite{hood2000,pinkse2000}. These
pioneering experiments open exciting new avenues for manipulating the motion of
neutral atoms, with the prospects for effective sub-Doppler cooling schemes
\cite{hechenblaikner98}, for controlled cavity-mediated long-range interactions
between atoms \cite{munstermann2000,gangl2000}, or for quantum information
processing \cite{Hemmerich,Zoller}.

The motion of a two--level atom in a free laser field is quite well understood
by now \cite{cohen92}. In a high--Q cavity, the confined radiation field 
has additional mechanical effects on the atomic center-of-mass (CM)
dynamics. The reason for this is the back-action of the atom on the field,
which becomes important in the regime of strong coupling between the atomic
dipole and the field.  Depending on its momentary position the atomic dipole
can drastically modify the radiation field which, in turn, exerts forces on the
moving atom. This results in a complex coupled dynamics of the atomic CM and
the field.

There is a fully quantum theory describing the cavity-induced mechanical
effects for intensities well below the single-photon level
\cite{hechenblaikner98}.  Analytical expressions have been derived for the
linear friction force and for momentum diffusion. For a large range of
parameters one finds strongly enhanced diffusion compared to free space, which
could be partly attributed to the quantum nature of the cavity field. 
Simulations based on these expressions are in good agreement with experimental
observations \cite{munstermann2000}. 

However, this theory is difficult to extend to higher intensities, where one
could expect better cooling and trapping due to reduced quantum diffusion. A
possible extension to higher photon numbers based on the standard approach for
laser cooling \cite{dalibard85} adapted to treat the situation in cavities was
recently presented by Doherty et al.\ \cite{doherty97,doherty2000}. The
computational effort, however, explodes with increasing photon numbers and the
approach is limited to calculate the dynamics in few photon fields.  Hence
there is a high demand for a model that allows for exploring the intermediate
regime between the quantum domain of subphotonic fields and classical laser
fields with large intensity. The interest is at least twofold. Firstly,
trapping and cooling of atoms in fields with intensities higher than the
single-photon level can lead to new phenomena. The forces get larger, hence the
potential wells involved are deeper, which is certainly interesting for further
applications. The ability of modeling and efficently simulating such systems is
one essential goal. Secondly, a simple model is needed for elucidating
fundamental features of the cavity-induced forces such as the enhanced or
suppressed momentum diffusion in comparison to free space.

In Sec.~II we develop a semiclassical model based on the Wigner representation
of both the motional atomic degrees of freedom and the cavity photon field,
including spontaneous atomic and cavity decay. Systematic truncation of the
corresponding time evolution equation at second order derivatives
yields a Fokker-Planck type equation which can be simulated by
classical stochastic differential equations. The essential feature of
this approach is the consistent treatment of the quantum noise of the
atom and cavity dynamics. The atomic momentum noise and the cavity
phase noise turn out to be correlated, which is the specific
ingredient of the atomic diffusion in a strongly coupled cavity. In
Sec.~III we use our simulation method to study the effects of atomic
localisation on the steady state properties of the system. We also
check the validity of our model by comparing the results with the
analytic ones in the weak excitation limit. In Sec.~IV we address the
question of the atomic trapping in the potential wells. We calculate
the characteristic trapping time, an important quantity in
experimental realisations. Finally, we summarise our results in
Sec.~V.

\section{Semiclassical model}

Let us consider a two--level atom interacting with a single mode of the
electromagnetic field in a weakly driven cavity. The driving field is supposed
to be monochromatic with frequency $\omega_p$ with a given detuning from the
mode frequency $\omega_C$ and with fixed effective amplitude $\eta$.
Dissipation in the system is due to cavity decay ($\kappa$) and spontaneous
emission from the atom into other than the cavity mode ($\gamma$). The atomic
dipole is strongly coupled to the field mode, i.e., the coupling constant $g$
is at least of the same order of magnitude as the relaxation rates $\kappa$ and
$\gamma$. The atom is allowed to move in the cavity under the effect of field
forces. The intracavity field is given by the mode function $f(x)$. For the
sake of simplicity this study will be restricted to one dimension but the
generalization of the model to three dimensions is straightforward.

We derive the semiclassical model by systematic approximations from
the full quantum master equation
\begin{equation}
\dot \rho = -\frac{i}{\hbar} \left[ H, \rho \right] + {\cal L}\rho \; .
\end{equation}
The coherent atom--field dynamics is considered in the electric dipole
and in the rotating wave approximation. In a frame rotating with the
pump frequency $\omega_p$, the Hamiltonian $H$ reads
\begin{equation}
H=\frac{p^2}{2 M} - \hbar \Delta_C a^\dagger a - \hbar \Delta_A
\sigma^+ \sigma^- 
-i\hbar g f(x) (\sigma^+ a - a^\dagger \sigma^-) - i\hbar \eta (a -
a^\dagger) \; , 
\end{equation}
where $\Delta_C=\omega_p-\omega_C$ and $\Delta_A=\omega_p-\omega_A$.
The operators $p$ and $x$ are
associated with the atomic momentum and position. The field and the atomic
dipole are described by the annihilation and creation operators $a$, 
$a^\dagger$, respectively by the lowering and raising operators $\sigma^-$,
$\sigma^+$. The damping terms are given by 
\begin{equation}
 {\cal L}\rho = -\kappa \left( a^\dagger a \rho + \rho  a^\dagger a - 2 a \rho
a^\dagger \right) -
\gamma  \left( \sigma^+ \sigma^- \rho + \rho  \sigma^+ \sigma^- - 2
\int_{-1}^1 N(u) \sigma^- e^{-i u x} \rho e^{i u x} \sigma^+ du
\right)\; ,
\end{equation}
where the last term includes the momentum recoil due to spontaneous
emission. 

It is possible to introduce an effective Hamiltonian of much simpler
form when the internal atomic dynamics can be adiabatically
eliminated. This is the case if the internal atomic variables
$\sigma^\pm$ evolve on a much more rapid time scale than the other
variables due to the large detuning $\Delta_A$ or due to the large
damping rate $\gamma$. In either case the population in the excited
atomic state is negligible (low saturation regime), and the atomic
dipole moment $\sigma^-$ can formally be replaced by
\begin{equation}
\sigma^- \approx - \frac{i\Delta_A + \gamma}{\Delta_A^2+\gamma^2} g f(x)
a \; .
\end{equation}
The dispersive and absorptive effect of the atom can be described by
the parameters
\begin{equation}
U_0= \frac{g^2 \Delta_A}{\Delta_A^2+\gamma^2}\; , \qquad 
\Gamma_0= \frac{g^2 \gamma}{\Delta_A^2+\gamma^2}\; ,
\end{equation}
respectively. The remaining atomic CM and field dynamics 
is governed by the adiabatic Hamiltonian
\begin{equation}
H_{\mbox{\tiny eff}}=\frac{p^2}{2 M} - 
\hbar \left(\Delta_C - U_0 f^2(x)\right)a^\dagger a 
- i\hbar \eta (a - a^\dagger) \; , 
\label{eq:heff}
\end{equation}
and the damping terms
\begin{equation}
{\cal L}_{\mbox{\tiny eff}}\rho = -\kappa \left( a^\dagger a \rho + 
    \rho a^\dagger a - 2 a \rho
   a^\dagger \right) -
\Gamma_0  \left(a^\dagger a f^2(x) \rho + \rho a^\dagger a f^2(x) - 
2 \int_{-1}^1 N(u) a f(x) e^{-i u x} \rho e^{i u x} f(x) a^\dagger du
\right)\; .
\label{eq:leff}
\end{equation}

Eqs.~(\ref{eq:heff}) and (\ref{eq:leff}) define an effective
quantum master equation that accounts for the coupled atomic CM motion
and field dynamics, and also for the environment induced losses. The
usual phase space method can be invoked to derive semiclassical
equations of motion \cite{GardinerZoller}, a technique which has been
successfully used for the study of laser cooling in free space, see e.g.\
Refs.~\cite{Javanainen,Castin,Petsas}.
We will omit the presentation of the detailed
calculation here and will concentrate on the main concepts only. The key for
deriving semiclassical equations from the quantum model is
the use of the combined Wigner function
$W(x,p,\alpha,\alpha^*)$ that represents the atomic motion {\it and}
the state of the cavity light field in the total phase space. The
Wigner transform of the quantum master equation leads to a partial
differential equation which is truncated at second-order derivatives
to obtain a Fokker--Planck-type equation (FPE). The validity of this
approach relies on two conditions.  First, the photon momentum $\hbar
k$ must be small compared to the momentum width $\Delta p$ of the
Wigner distribution, i.e., a single-photon absorption or emission will
not change the momentum distribution considerably. Accordingly, the
powers of the small parameter $\hbar k/\Delta p \ll 1$ introduce a
hierarchy in the different orders, which justifies the truncation
procedure. 
Besides this condition which is well known from semiclassical treatments of
free-space laser cooling, there is a second
one concerning the quantized field state. That is, the second-order
derivative $\frac{\partial^2}{\partial\alpha\partial\alpha^*}$ must be
neglected compared to $|\alpha|^2$, in order to drop a term containing
third-order derivatives in momentum and field variables.
A sufficient condition is to
consider fields close to coherent states with mean photon numbers
higher than one. This step makes the model ``semiclassical'' in the description
of the cavity field as well.

The FPE for the combined atom-field Wigner function reads
\bea
\frac{d}{dt} W &=& \left[i\left(U_0 f^2(x)-\Delta_C\right) \left(
\frac{\partial}{\partial\alpha} \alpha -
\frac{\partial}{\partial\alpha^*}\alpha^* \right) - \eta  \left(
\frac{\partial}{\partial\alpha}+\frac{\partial}{\partial\alpha^*}
\right) \right] W\nonumber\\
& & 
+ \left[\left(\kappa + \Gamma_0 f^2(x)\right) \left(
\frac{\partial}{\partial\alpha} \alpha +
\frac{\partial}{\partial\alpha^*}\alpha^* +
\frac{\partial^2}{\partial\alpha\partial\alpha^*} \right) \right] W 
\nonumber\\
& &
+ \left[- \frac{p}{M} \frac{\partial}{\partial x} 
+ U_0 \left(|\alpha|^2-\frac{1}{2} \right) \hbar \nabla f^2(x)
\frac{\partial}{\partial p}\right] W \nonumber\\
& & 
+ \Gamma_0\left(|\alpha|^2-\frac{1}{2} \right)
\left( (\hbar \nabla f(x))^2 + \hbar^2 k^2\bar{u^2} f^2(x)\right)
\frac{\partial^2}{\partial p^2}W\nonumber\\
& & 
+ i \Gamma_0 \hbar f(x) \nabla f(x) \left(
\frac{\partial}{\partial\alpha} \alpha -
\frac{\partial}{\partial\alpha^*}\alpha^*
\right)\frac{\partial}{\partial p} W,
\label{FPE}
\eea
where the first term is the coherent evolution of the cavity field, the second
term is the decay and diffusion of the cavity field due to cavity decay and
scattering of photons out of the cavity by the atom, the third term is the
conservative atomic motion, and the forth term is the momentum diffusion due to
scattering of photons by the atom. Note that this latter term can become
negative, which reflects the break-down of the semiclassical model at very weak
intracavity field intensities. Additionally, we find a rather unintuitive fifth
term which gives rise to correlated momentum and cavity field noise. We will
discuss this in more detail later. In (\ref{FPE}) we introduced the
abbreviation $\bar{u^2} = \int N(u) u^2 du$,  which we will set to $0.4$ in the
following, valid for circularly polarised light.

In the case of positive Wigner functions $W$, the FPE can be solved by
Monte-Carlo simulations of the corresponding stochastic differential equations
(SDE) \cite{Gardiner}
\bea
dx & = & \frac{p}{M} dt, \nonumber \\
dp & = & -\hbar U_0 \left( \alpha_r^2+\alpha_i^2 -\frac{1}{2}\right)
         \nabla f^2(x) dt + dP, \nonumber \\
d\alpha_r & = & -\eta dt + \left(U_0f^2(x)-\Delta_C\right)\alpha_i dt
                - \left(\kappa + \Gamma_0 f^2(x)\right)\alpha_r dt + dA_r, 
		\nonumber \\
d\alpha_i & = & - \left(U_0f^2(x)-\Delta_C\right)\alpha_r dt
                - \left(\kappa + \Gamma_0 f^2(x)\right)\alpha_i dt + dA_i,
\label{eq:SDE}		
\eea
where we decomposed the cavity field into its real and imaginary parts, $\alpha
= \alpha_r + i\alpha_i$. We thus have reduced the solution of the FPE to a set
of four SDE which is easily tractable by numerical integration. Another major
advantage of this approach as compared to previously used methods
\cite{horak97,doherty97} is that the numerical effort is independent of the
photon number. Thus our method is suitable for exploring the
transition from the quantum fields with very low photon numbers all
the way to the high-intensity classical regime \cite{Vuletic}. 
Note however that due
to the adiabatic elimination of the atomic excited state the method
presented here only works for small atomic saturation, which is not
guaranteed for some experiments even in the low photon regime.

The most intricate part in a numerical simulation of the SDE is the correct
treatment of the noise terms $dP$, $dA_r$, and $dA_i$ since these are
correlated, as already mentioned above. For a discussion of the correlated
diffusion, it is convenient to introduce the field amplitude noise
$dA_\|$ and the field phase noise $dA_\perp$ by a rotation of $dA_r$ and
$dA_i$,
\bea
dA_\| & = & \frac{\alpha_r}{|\alpha|} dA_r + \frac{\alpha_i}{|\alpha|} dA_i,
   \nonumber \\
dA_\perp & = & -\frac{\alpha_i}{|\alpha|} dA_r + \frac{\alpha_r}{|\alpha|} dA_i.
\eea
The diffusion matrix $\bf D$ in this basis then reads
\be
{\bf D}\, dt = 
\langle 
   \left(
   \begin{array}{c}
   dA_\| \\
   dA_\perp \\
   dP
   \end{array}
   \right) 
   \left(dA_\|,\, dA_\perp,\, dP \right) 
\rangle =
\left(
\begin{array}{ccc}
d_1 & 0 & 0 \\
0 & d_1 & d_3 \\
0 & d_3 & d_2
\end{array}
\right) dt
\label{eq:diffmatrix}
\ee
where
\bea
d_1 & = & \frac{1}{2}\left(\kappa+\Gamma_0 f^2(x) \right), \nonumber \\
d_2 & = & 2\Gamma_0\left(|\alpha|^2-\frac{1}{2} \right)
   \left( (\hbar \nabla f(x))^2 + \hbar^2 k^2\bar{u^2} f^2(x)\right),
   \nonumber \\
d_3 & = & \Gamma_0 |\alpha| \hbar f(x) \nabla f(x).
\label{eq:diffcomp}
\eea
From this we see that the amplitude noise is in fact independent, but the phase
noise and the momentum noise are correlated. We interpret this in the following
way. In a free space description it is well-known \cite{cohen92} that a
two-level atom gives rise to photon redistribution between the two
counterpropagating travelling waves which form the standing wave.\footnote{Note
that, strictly speaking, the decomposition of a single standing wave mode of a
resonator into two travelling waves does not make sense. However, this provides
an intuitive picture of the physical processes and is only used here in this
sense.}  The incoherent part of this redistribution is proportional
to $\Gamma_0$ and accounts for one part of the atomic momentum diffusion. In
standard weak-field Doppler cooling \cite{cohen92}, the backaction of this
scattering process on the light field is neglected as this is assumed to be
fixed by the driving laser. However, in the strong coupling limit of an optical
cavity this backaction must be taken into account. Since such a spontaneous
backscattering process does not change the photon number in the standing wave
mode, there is no corresponding intensity fluctuation but, depending on the
atomic position, the phase of the cavity field is changed. This leads to the
correlated momentum and phase noise represented by the nondiagonal elements of
the diffusion matrix ${\bf D}$.

Note that the cavity field noise $d_1$, Eqs.~(\ref{eq:diffmatrix}) and 
(\ref{eq:diffcomp}) is independent of the field amplitude $\alpha$. Hence, the
noise terms $dA_{r,i}$ in the SDE (\ref{eq:SDE}) play an important role in the
system dynamics for small $\alpha$. For large field amplitudes, $d_1$ is small
compared to the diffusion matrix elements $d_2$ and $d_3$ and can be neglected.
In this limit we recover the limit of atomic motion in a dynamically varying
classical light field. If additionally the atomic saturation is reduced by
increasing the detuning $\Delta_a$, also the atomic momentum diffusion $dP$
becomes negligible compared to the coherent dynamics. We then obtain the
classical equations of motion of a cassical dipole in a high-finesse cavity
\cite{horak97,hechenblaikner98}. This limit has been recently suggested as a
possible scheme for laser cooling of molecules, where one tries to avoid
spontaneous atomic transitions to uncoupled molecular states \cite{Vuletic}.

Let us now compare our method with previous semiclassical simulations used
e.g.\ by Doherty et al.\ \cite{doherty97,doherty2000}. In that approach, the
full master equation is numerically solved up to first order in velocity
for any position in space, which gives the local dipole and friction forces.
Additionally, the momentum diffusion coefficient is obtained by calculating
the force autocorrelation function using the quantum regression theorem. Hence,
no semiclassical assumption for the cavity field is imposed, and arbitrary
atomic saturation is taken into account. In this treatment the numerical effort
increases drastically with increasing photon number and becomes exceedingly high
in the case of several atoms or cavity modes, as for instance in a ring
cavity. In contrast, our semiclassical method is easy to implement also in these
more complex situations. In addition, the explicit dynamical evolution of the
cavity field in our case allows for arbitrary atomic velocities and provides new
insight into the coupled atom-cavity dynamics, such as the correlated noise
sources discussed above.

\section{Temperatures and localisation effects}

We will now turn to the discussion of some specific numerical examples
and compare these with previous analytic expressions in the weak
excitation limit \cite{horak97,hechenblaikner98}. The validity of the
latter is restricted to parameter regimes with mean intensity well
below the single-photon level. Nevertheless, the analytic solution is
expected to be a good approximation for higher photon numbers as long
as the atomic saturation is very low.  As an example, we will consider
a standing wave with mode function $f(x)=\cos(k x)$. First, we will
study the cavity-induced effects on the ``temperature'' (the
time-averaged kinetic energy) of a single atom in the one-dimensional optical
lattice. Next, the spatial distribution of the atomic position is
considered, and the effect of localisation on the temperature will be
demonstrated.

\begin{figure}
\infigtwo{7cm}{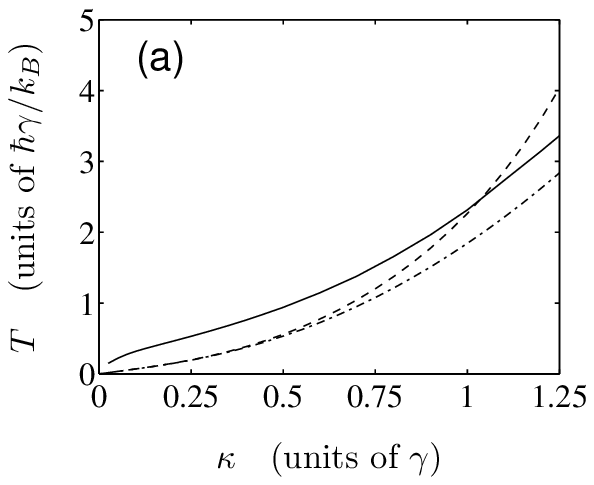}{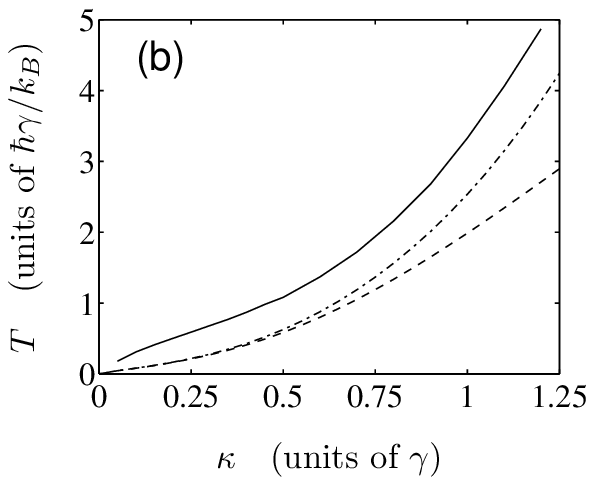}
\caption{Steady-state temperature versus cavity decay rate for (a)
$\Delta_A=20\gamma$, $\Delta_C=0$ and (b) $\Delta_A=-20\gamma$,
$\Delta_C=U_0=0.312\gamma$. Solid curves show the numerical results obtained
from our semiclassical model. The other curves correspond to the analytical
results in the weak-driving limit without (dashed) and with (dash-dotted)
adiabatic elimination of the atomic excited state. The other parameters are
$\eta=3\kappa$, $g=2.5\gamma$, $\gamma=2\pi\times3$MHz (rubidium atoms).
}
\label{fig:kappa}
\end{figure}

The steady-state temperature as a function of the cavity decay rate is
plotted for the two parameter regimes where the cavity-mediated force
on the atom is known to give rise to cooling, that is, for
$\Delta_A>0$, $\Delta_C=0$ in Fig.~\ref{fig:kappa}(a) and for
$\Delta_A<0$, $\Delta_C=U_0$ in Fig.~\ref{fig:kappa}(b). For the
numerical simulations we keep the ratio of pump strength to cavity
decay rate constant, $\eta/\kappa=3$, such that the cavity contains on
average approximately nine photons for all parameter
configurations. The variation of the temperature can then be
associated with the effect of the cavity. The cavity-induced forces
are expected to be important in the range $\kappa \approx U_0 = 0.312
\gamma$, while the limit of $\kappa, \eta \rightarrow \infty$
corresponds to a free-space radiation field. For reference, the
analytical results of Ref.~\cite{hechenblaikner98} have been used to
obtain two curves for the temperature, based on the total friction
force and, respectively, on the cavity-induced friction force
only. The difference between these analytic results is thus due to the
standard Doppler force which is heating for $\Delta_A>0$ and cooling
for $\Delta_A<0$. As expected, the numerical results are closer to
the analytic curves without the Doppler force, because the
adiabatic elimination of the atomic excited state excludes the Doppler force
from our semiclassical model.  This is a good
approximation since the Doppler shift $k v$ is much smaller than the
atomic detuning $\Delta_A$. However, for increasing cavity linewidth,
$\kappa \gg U_0$, the cavity-induced friction force as well as the
diffusion vanish. It is no longer justified to neglect the
Doppler force in this limit, indicated by the bifurcation point of the
two analytic curves. Nevertheless, even within the limit of validity
we find a systematic deviation of the numerical and the analytic
results: the simulated temperatures are always higher. In the
following we show that the difference can be attributed to atomic
localisation in the potential wells, which is treated correctly within
the semiclassical model but not contained in the weak-driving
limit of the analytic formulas.

\begin{figure}
\infigtwon{7cm}{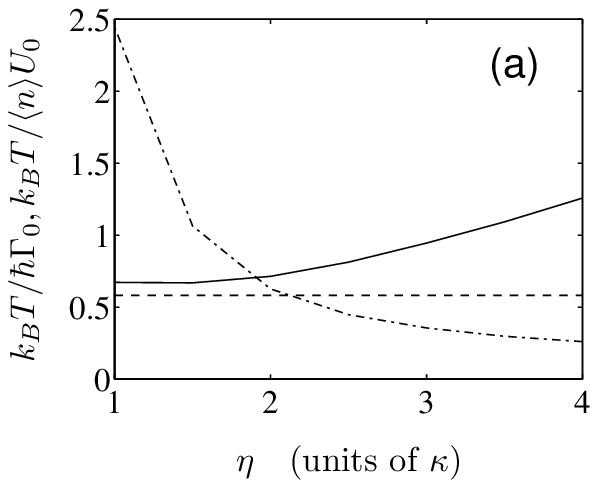}{6.8cm}{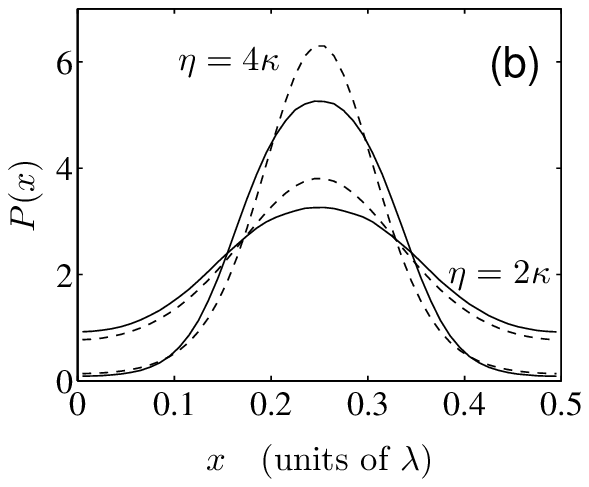}
\caption{Effects of localisation for different pump strengths. (a) Steady-state
temperature $T$ versus $\eta$ (solid line), temperature obtained from the
analytic solution in the weak-driving limit (dashed), and ratio of $T$ and the
mean optical potential $U_0\langle n\rangle$. (b) Atomic position distributions
within a potential well for $\eta=2\kappa$ and $\eta=4\kappa$ (solid lines), 
respectively, as compared with thermal distributions (dashed) of the same
temperature and the same mean potential. For this plots $\kappa=\gamma/2$ and
the other parameters are as in Fig.\ \protect\ref{fig:kappa}(a).}
\label{fig:eta}
\end{figure}

In order to study the effects of localisation in more detail, we show
in Fig.~\ref{fig:eta}(a) the dependence of the steady-state
temperature on the pump strength when all other parameters are kept
constant. We see that for decreasing pump strength (decreasing photon
number) the temperature approaches the analytic result of the
weak-excitation limit. Note that for very small photon numbers of 1 or
2 our semiclassical model is no longer valid and the numerical results
exhibit a residual deviation from the analytical ones. (One of the
numerical signatures of the limited validity of our model in this
parameter regime is the occurence of a negative eigenvalue of the
diffusion matrix $\bf D$, Eq.~(\ref{eq:diffmatrix}), in single
particle trajectories.)

Figure~\ref{fig:eta}(a) also shows the ratio of the steady-state
temperature and the time-averaged optical potantial depth given by the optical
potential times the average photon number $U_0\langle n\rangle$, with
$n=a^\dagger a$. In the weak driving limit, this ratio is much larger
than one, indicating a nearly flat atomic distribution in position
space. For stronger pumping, the ratio is continuously decreasing and
drops well below unity corresponding to strong localisation. This
indicates a strong {\em local} cooling force, that is, even a particle
which is trapped within a single potential well experiences a friction
force. Hence, in order to achieve good trapping of particles in the
cavity QED domain, one should go to higher photon numbers than those
which are currently used in experiments, despite the corresponding
higher steady-state temperatures.  We will return to the discussion of
trapping times in the following section.

Finally, we show in Fig.~\ref{fig:eta}(b) the averaged spatial
distribution of a particle for two different values of the pump
strength.  This again shows the stronger localisation for increasing
$\eta$. Moreover, the dashed lines give the spatial distribution of a
particle obtained from a thermal distribution with the same mean
temperature and a sinusoidal potential of the same mean depth, that
is, \be P_{\mbox{\tiny therm}}(x) \propto \exp\left(-\frac{U_0\langle
n\rangle\cos^2(kx)}{k_B T}\right).  \ee Note that the numerically
obtained distribution strongly deviates from a thermal one due to
dynamical effects and the corresponding correlations between atomic
degrees of freedom and the cavity field.

\section{Trapping times}

For cavity QED experiments, especially in view of potential applications in
quantum information processing, it is an important issue how long
the particles can be trapped in the potential wells. Depending on the
initial condition, there are different ways to discuss the trapping
time of an atom in a well.

\begin{figure}
\infig{7cm}{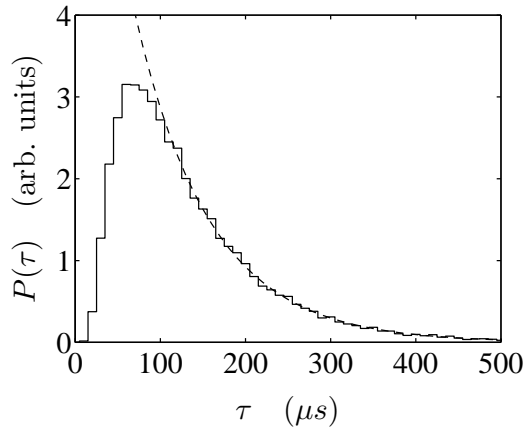}
\caption{Probability distribution $P(\tau)$ for the times at which an
atom initially in the ground state of a potential well leaves this well. The
solid line is a histogram obtained numerically and the dashed line is an
exponential fit.
The parameters are those of Fig.\ \protect\ref{fig:eta}(b) and $\eta=2\kappa$.}
\label{fig:purestate}
\end{figure}

At time $t=0$ the atom can be assumed in the ground state of the potential
well. The potential is approximately harmonic so the Wigner function of the
ground state of a harmonic oscillator can be used as initial distribution in
the semiclassical Monte-Carlo simulations.  For a large statistical ensemble we
then record the times $\tau$ when the atoms leave the initial potential well. A
typical example for the resulting distribution $P(\tau)$ is presented in figure
\ref{fig:purestate}. We find a low probability that an atom leaves the
potential well at short times since it takes a while until the particle is
sufficiently heated by momentum diffusion. For larger times the distribution
becomes exponentially decaying
$P(\tau)\approx\exp(-\tau/T_{\mbox{\small trap}})$, e.g., for the given
parameters and $\tau>100\mu s$ a fit yields the trapping time
$T_{\mbox{\small trap}}=88.3\mu s$. Note that this initial state is in fact
realised in recent experiments \cite{pinkse2000} where atoms from a 
magneto-optical trap are velocity selected through the narrow cavity entrance
slits before they interact with the optical field, and thus the atoms are
effectively prepared in their motional ground state as an initial condition.

\begin{figure}
\infig{7cm}{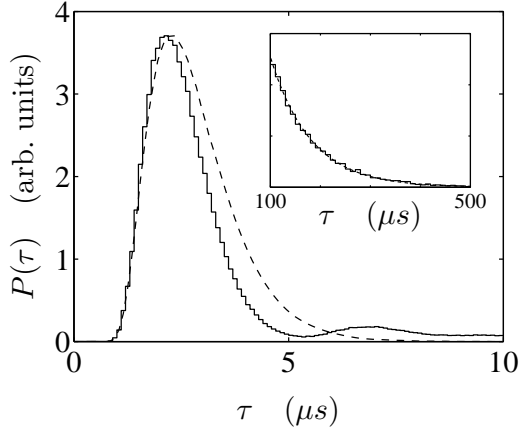}
\caption{Distribution of flight times which an atom spends within single
potential wells when it is observed for a very long time. The solid line
shows the numerical results, the dashed line is the result for a thermal
ensemble of atoms in a static potential for comparison.
The inset shows an exponential fit (dashed curve) for long times $\tau$.
The parameters are the same as in Fig.\ \protect\ref{fig:purestate}.}
\label{fig:traptime}
\end{figure}

Alternatively one could consider a thermalised atom as initial condition. In
this case we simulate a single atom trajectory for a very long period and
record the times $\tau$ which the atom spends within a potential well. In the
following we will call $\tau$ a ``flight time'' in order to distinguish it from
the trapping time $T_{\mbox{\small trap}}$ as defined before. We then obtain
a distribution for short and long flight times as presented in Figure
\ref{fig:traptime}. 

As a help for the interpretation of this curve, let us first
discuss the corresponding distribution obtained for a thermal ensemble
with the same temperature and in a static (conservative) potential of
the same depth $\langle n\rangle U_0$. In this case there is no
momentum diffusion and hence an atom which is trapped initially will
stay within the same potential well for arbitrarily long
times. Therefore only hot atoms with a total energy larger than the
optical potential depth will contribute to the distribution function
of flight times which is shown by the dashed curve in
Fig.~\ref{fig:traptime}. This distribution function shows no flight times
$\tau<1\mu s$ that would correspond to very fast atoms, a single maximum at
$\tau\approx 2.3\mu s$, and a continuously decreasing probability for longer
flight times.

Taking the full cavity dynamics into account has two main
effects. First, because of momentum diffusion every atom undergoes a
random walk in momentum space and thus will leave a given potential
well after a finite time. Second, if an atom enters a potential well
with a not too high velocity, it will be slowed down due to the
cavity-mediated friction force and thus can be trapped for a longer
time. This effect manifests itself in the decrease of the distribution
function for times $2\mu s<\tau<6\mu s$ compared to the conservative
motion limit. However, in general a particle which just has been
trapped will have an energy only slightly below the potential maximum
and therefore a relatively high probability of being ejected from the
well within a short time. This leads to the second maximum of the
distribution function at $\tau\approx 7\mu s$.

Finally, for long times the exponential distribution of
the trapping times is detected again, with 
$T_{\mbox{\small trap}}=80.5\mu s$, in reasonable agreement with the
previous result in
Fig.~\ref{fig:purestate}. Note, however, that the two definitions for the
trapping time $T_{\mbox{\small trap}}$ are qualitatively different: in 
Fig.~\ref{fig:purestate} the distribution is obtained from an ensemble of
statistically independent atoms, whereas in Fig.~\ref{fig:traptime} only a
single atom is followed. Hence in the latter case the distribution exhibits a
certain memory of the particle history. For example, a very fast atom will
travel over many potential wells and thus will produce many entries in the
distribution at short flight times until it gets slowed down and trapped. 
Because of this the large maximum at times $1\mu s<\tau<5\mu s$ in 
Fig.~\ref{fig:traptime} only corresponds to about 22\% of the total time for
which the atom was observed. This also agrees well with the fraction of 20.4\%
of atoms with energies above the potential barrier obtained from a thermal
ensemble, which further supports our interpretation of the distribution function
$P(\tau)$ in terms of trapped and untrapped atoms.

\begin{figure}
\infig{7cm}{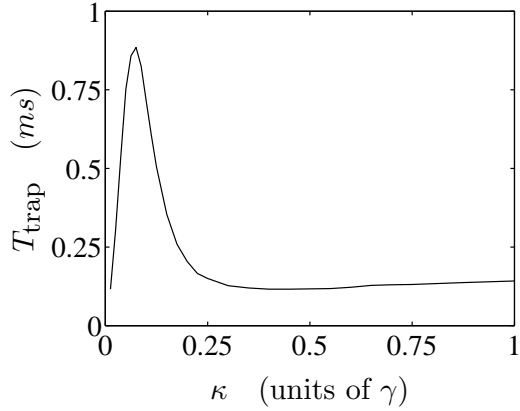}
\caption{Trapping time $T_{\mbox{\small trap}}$ versus $\kappa$. 
The parameters for this plot are those of
Fig.~\protect\ref{fig:kappa}(a).}
\label{fig:tt_kappa}
\end{figure}

\begin{figure}
\infig{7cm}{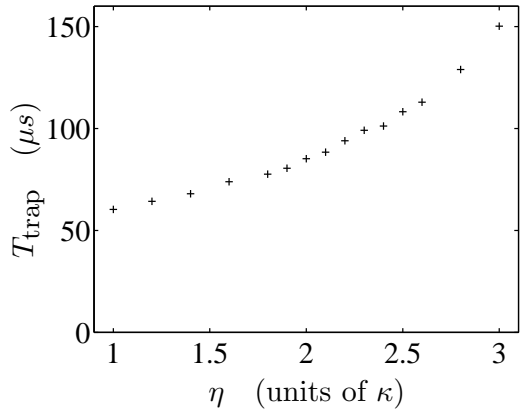}
\caption{Trapping time $T_{\mbox{\small trap}}$ versus pump strength $\eta$
with $\kappa=\gamma/2$. The
parameters are the same as in Fig.\ \protect\ref{fig:eta}. }
\label{fig:tt_photon}
\end{figure}

The above outlined method for calculating the characteristic trapping time
$T_{\mbox{\small trap}}$ has been carried out for different parameter settings.
The trapping time as a function of $\kappa$ and of the pump strength $\eta$ is
shown in Figures \ref{fig:tt_kappa} and \ref{fig:tt_photon}, respectively. 

Figure \ref{fig:tt_kappa} is similar to the analysis of
Fig.~\ref{fig:kappa}(a), where the photon number is kept approximately
constant by a fixed ratio of $\eta/\kappa=3$, and the cavity-induced
effects are modified by varying the cavity decay rate
$\kappa$. Depending on the value of $\kappa$ we find three very
distinct regimes for the behaviour of the trapping time. First, there
is a strong increase of the trapping time for very small decay
rates. This is due to the strong coupling of the photon number to the
atomic position in this limit: already a small elongation of an atom
from the field node is sufficient to shift the cavity out of resonance
and to drastically reduce the photon number. Thus, the atom
dynamically reduces the trapping potential and subsequently can easily
leave the potential well. Hence the trapping times are short in this
limit. For $\kappa>U_0$ the photon number can at most be changed by a
factor of two by the atom and the behaviour of the trapping time is no
longer dominated by this dynamical effect. In this regime the photon
number and consequently the potential depth are approximately
constant, but the temperature of the atom increases with increasing
$\kappa$ as we have seen in Fig.~\ref{fig:kappa}(a). Thus the atomic
localisation decreases, which in turn gives rise to a decrease of the
trapping time. Finally, we find again a slight increase of the
trapping time when $\kappa$ is so large that the temperature is of the
order of or larger than the optical potential depth. An increase of
$\kappa$ then still increases the temperature and thus the fraction of
untrapped particles. However, our definition of the trapping time as
the exponential decay of the flight time distribution for very long
times (Fig.~\ref{fig:traptime}) relies on the {\em trapped} particles
only and does not depend on the fraction of untrapped atoms. The
trapping time is then dominated by the time scale of the friction
force and of the momentum diffusion, which becomes longer for large
increasing $\kappa$. This is due to approaching the classical
potential limit for $\kappa\rightarrow\infty$ with a fixed ratio
$\eta/\kappa$. This explains the final increase of the trapping
time in Fig.~\ref{fig:tt_kappa}.

In Figure \ref{fig:tt_photon} the photon number is varied by $\eta$, while
the other parameters are kept constant. Though the temperature gets
higher with increasing photon numbers (see
Fig.~\ref{fig:eta}), the resulting trapping times exhibit a
considerable increase owing to the strong deepening of the potential
wells. This result suggests clearly the benefit of working in an
intermediate photon number regime for implementing coherent
manipulation with neutral atoms in optical lattices.

\section{Conclusions}

We have derived semiclassical stochastic differential equations to simulate the
coupled dynamics of an atom moving in a single cavity mode in the limit of
strong nonresonant interaction between the field and the atomic dipole.
Starting from the basic quantum equations we performed systematic
approximations such that the model accounts consistently for the quantum noise
properties of the system. A nontrivial correlation between the field phase
noise and the momentum noise has been revealed and can be interpreted in terms
of simple physical processes. 

As an example for our method we have studied the atomic motion in a one
dimensional optical lattice sustained by the strongly coupled cavity field. In
the appropriate limits the calculated steady-state temperatures are in good
agreement with the results of previous calculations. Deviations can be
attributed to atomic localisation within the potential wells, which is
consistently described in our semiclassical simulation. We found that for higher
intracavity photon numbers the characteristic trapping times can be
significantly increased up to milliseconds. In this case the potential wells
get deeper and  diffusion and friction are enhanced in a way that the
temperature only slowly increases. Hence the steady state temperature can get
significantly smaller than the well depth. Alternatively by simultaneously
increasing the atom-field detuning, we can keep the effective potential the
same but reduce diffusion to achieve longer storage times. As the calculational
effort only weakly depends on the photon number we can nicely study the
transition from a quantum to a classical field.

Let us mention here that the model can easily be generalised to 3D, more cavity
modes and several atoms simultaneously in the cavity. Hence, this semiclassical
approach offers a tool for a tractable numerical simulation of the
dynamics of these complex systems. These simulations could then be used to
check the performance of atom cavity systems for quantum computing and quantum
information processing.

\acknowledgments

We thank P.\ W.\ H.\ Pinkse, T.\ Fischer, P.\ Maunz, and G.\ Rempe for
stimulating discussions. This work was supported by the Austrian
Science Foundation FWF (Project P13435). P.~D.\ acknowledges the
financial support by the National Scientific Fund of Hungary under
contracts No. T023777 and F032341.

\end{document}